\documentclass[prb,superscriptaddress,amsmath,amssymb,twocolumn,nofootinbib]{revtex4-2}

\usepackage[normalem]{ulem}
\usepackage[utf8]{inputenc}
\usepackage{textcomp}
\usepackage{caption}
\usepackage{lipsum}
\usepackage{graphicx}
\usepackage{times}
\usepackage{mathptmx}
\usepackage{tipa}
\usepackage{subcaption}
\usepackage{booktabs}
\usepackage{latexsym}
\usepackage{xcolor}
\usepackage{svg}
\usepackage{makecell}
\usepackage[verbose]{placeins}
\usepackage{setspace}
\usepackage{tikz}
\usepackage{hyperref}
\usepackage{comment}

\usepackage[textsize=scriptsize, color=orange!60]{todonotes}

\usetikzlibrary{backgrounds,fit,decorations.pathreplacing,calc}

\usepackage[sectionbib]{bibunits}
\defaultbibliographystyle{apsrev4-1}
\defaultbibliography{references}

\begin{document}

\title{Standardized test of many-body coherence in gate-based quantum platforms}

\author{Yi Teng}
\affiliation{TCM Group, Cavendish Laboratory, University of Cambridge, Cambridge CB3 0HE, UK}

\author{Orazio Scarlatella}
\affiliation{TCM Group, Cavendish Laboratory, University of Cambridge, Cambridge CB3 0HE, UK}

\author{Shiyu Zhou}
\affiliation{Perimeter Institute for Theoretical Physics, Waterloo, Ontario, Canada N2L 2Y5}

\author{Armin Rahmani}
\affiliation{Department of Physics and Astronomy and Advanced Materials Science and Engineering Center, Western Washington University, Bellingham, Washington 98225, USA}

\author{Claudio Chamon}
\affiliation{Department of Physics, Boston University, Boston, MA, 02215, USA}

\author{Claudio Castelnovo}
\address{TCM Group, Cavendish Laboratory, University of Cambridge, Cambridge CB3 0HE, UK}

\begin{abstract}
Quantum coherence is a crucial resource in achieving quantum advantage over classical information processing, and more generally developing new quantum technologies. While its effects are observable in current quantum platforms, there are no standardized tools for systematically measuring and quantifying multi-qubit coherence across different gate-based quantum hardware. In this work, we propose a method to define a many-body quantum coherence length scale using anyon interference effects in a spin-chain setup, which effectively mirrors the problem of a quantum particle on a ring, with or without flux through it. We propose using the maximum length of the ring for which the presence or absence of flux can be clearly discerned, as a simple measure of the many-body \emph{quantum coherence grade} (Q-grade) in a given quantum hardware. We demonstrate how this approach can be implemented on gate-based quantum platforms to estimate and compare the quantum coherence of current devices, such as those from Google, IBM, IonQ, IQM, and Quantinuum that we considered here.
This work aims to contribute to the creation of a live Web interface where the latest developments and advancements can be demonstrated, and progress in quantum coherence resources tracked over time. Establishing such a standardized quantum test would enable monitoring the growth of quantum coherence in gate-based quantum platforms, in a spirit similar to Moore's law. 
\end{abstract}

\maketitle
%
%

\begin{bibunit}

\section{Introduction
\label{sec:intro}}
Gate-based quantum platforms have steadily and impressively advanced over the past few years~\cite{google_quantum_ai,ibm,ionq,iqm,quantinuum}. Improvements in single-qubit coherence times and gate fidelity in the hardware have brought us to a stage in which simple and short quantum computations can already be demonstrated~\cite{Arute_2019,Kim_2023,Zhao_2023,Iqbal_2024}. Incorporation of quantum error correction~\cite{Calderbank_1996, Gottesman_1997, Knill_1997} promises to further stretch the register sizes and computational times possible with gate-based platforms. 

A key resource in the quest to achieve quantum advantage in new quantum technologies is quantum coherence. However, to date simple and systematic tools to measure and quantify multiple-qubit coherence across different types of gate-based hardware, and most importantly through time as these platforms evolve, are not yet 
available. 
This achievement would be of particular importance to track a possible Moore's law of quantum coherence and predict future capabilities to plan efforts and investments. 

The purpose of this paper is to propose such a standardization using an interferometry-based test~\cite{Zhou_2024}, guided by the following criteria. 
First, the test should be based on a minimalist and archetypal dynamical problem in quantum physics. Second, it must translate into a standard circuit description that implements Trotterized evolution~\cite{Trotter_1959,Lloyd_1996}, mimicking the dynamics of the chosen problem. Third, the test should involve running \emph{the same circuit} with two different initial states, each corresponding to a distinct physical realization of the underlying quantum physics problem, whose outputs can be compared and contrasted.

The minimalist example that we chose is inspired by a `particle on a ring' -- a simple problem physicists learn in early studies of quantum mechanics. We choose a discretized version of the problem, i.e., a tight-binding version of it. 
The periodic boundary conditions geometry allows one to place a `flux' through the ring, and therefore compare the time-evolution of the wavefunction of a particle launched at one site, in the presence or absence of flux. 
Crucially, we set the problem up in an emergent way, in order to satisfy our third criterion. Drawing inspiration from the celebrated Kitaev's toric code~\cite{Kitaev_2003}, our particle and flux are emergent topological quasiparticles, 
with non-trivial exchange statistics~\cite{Wen_2007}. 
We can then turn the flux on or off in our ring geometry by introducing an effective `magnetic charge' -- a vison -- at the center of the ring by an appropriate choice of initial state. The particle is encoded in a ferromagnetic domain wall -- a spinon -- whose Hamiltonian dynamics can be written in simple Ising-chain form, 
which we then translate into a conventional Trotterized circuit, as required by our second criterion. The family of circuits thus only depends on the size of the ring, $L$. The many-body \emph{quantum coherence grade} (Q-grade) of a given quantum hardware 
is then the largest $L$ for which one can sufficiently distinguish the two situations -- flux or no flux -- by contrasting the probabilities that the particle arrives at the diametrically opposite site to the one it is launched from, as explained below.

The Q-grade with which a given hardware passes this standard test should grow 
as technology evolves and improves. We stress that the test defines a standard circuit and standard encoding for the physical problem that motivates it; of course, one can change the encoding to more efficiently simulate the same physical problem, but one should not confound presenting a physically motivated standard test of a quantum hardware with solving the problem at hand more cleverly (which is ultimately known and trivial, and can be solved classically). In short, the test is not meant to grade the skills of `quantum software programmers'; it is designed to log and monitor the advances in quantum hardware development.
%
%

\section{Model
\label{sec:model}}
\begin{figure}[!h]
    \centering
    \includegraphics[width=.4\textwidth]{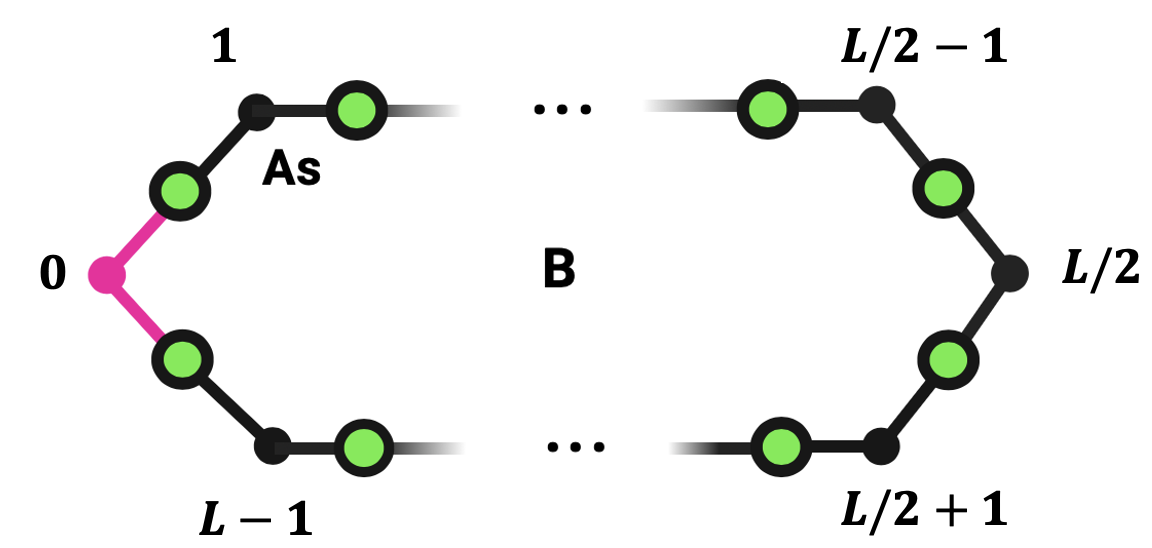}
      \caption{Schematic illustration of the system. The green-filled large dots represent the qubits, which interact ferromagnetically across all black links, and antiferromagnetically across the pink link (twisted boundary conditions). The small solid dots on the vertices denote the spinon sites (labelled clockwise as $0,\dots, L-1$). The ring can be occupied by a vison ($B=-1$), or empty ($B=+1$).} 
    \label{fig:system}
\end{figure}
While the quasiparticle content and fractional statistics of our chosen model is best understood in light of Kitaev's toric code~\cite{Kitaev_2003}, the model is in fact most simply stated as a ferromagnetic Ising ring (nearest-neighbour interaction $J=1$, our reference energy) with twisted boundary conditions (namely, one bond has energy $-J$; bond $s=0$ in Fig.~\ref{fig:system} and Eq.~\eqref{eq:Ham}). This enforces a strictly odd number of domain walls (defined as bonds across which the interaction energy is not minimised), with one domain wall in the entire system having lowest energy. The domain walls play the role of particles in our context, equivalent to `e' charges or spinons in Kitaev's toric code. 

If we implement the interactions between the $z$ components of qubits, one finds that all terms commute (trivially) with one another and (importantly) with the product of all $x$ components of the qubits around the entire ring. The eigenvalues of such product operator, $B = \pm 1$, are therefore good quantum numbers of the system. This encodes our flux: when $B=-1$, the motion of a domain wall around the ring causes an overall change of sign in the wavefunction of the system, whereas when $B=+1$ the system returns to the very same state. $B = \pm 1$ corresponds to the absence / presence of an `m' charge or vison in Kitaev's toric code. 

Further details of the model can be found in SM~Note~\ref{sm_sec:model}, and a pictorial illustration is provided in Fig.~\ref{fig:system}. 
In order to give dynamics to our system, we introduce a transverse field $0 < \Gamma \ll J$, 
\begin{equation}
H = J \, A_0 - J \sum_{s \neq 0} A_s 
- {\Gamma} \sum_i\sigma_i^x 
\, ,
\label{eq:Ham}
\end{equation}
where $A_s = \prod_{i \in s} \sigma_i^z$, with the leading effect of providing a hopping amplitude for the spinons. The transverse field also produces matrix elements between spinon number sectors, via spinon pair creation and annihilation events; however, these are energetically suppressed and play a lesser role in the discussion below, given the choice of $\Gamma \ll J$ (see in Fig.~\ref{fig:lindblad}(b)). 

In real devices, decoherence is unavoidable and coherent quantum mechanical time evolution is progressively curtailed by noise~\cite{Preskill_2018}. The effect can be modelled for instance by coupling the system to a Markovian isotropic and uniform bath \textit{\`a la} Lindblad~\cite{Lindblad_1976} (see SM~Note~\ref{sm_sec:dephasing}): 
\begin{equation}
\dot{\rho} 
= 
- i [H,\rho] 
+ \gamma \sum_i \left(
  \sigma^x_i \rho \sigma^x_i + \sigma^y_i \rho \sigma^y_i + \sigma^z_i \rho \sigma^z_i - 3 \rho
\right) 
\, ,
\label{eq:lindblad}
\end{equation}
where $\rho$ is the density matrix describing the system, and $\gamma$ parametrises the strength of the noise. 
%
%

\section{Protocol
\label{sec:protocol}}
Noise makes it challenging to observe the effects of quantum coherence, and distinguish them from , e.g., spurious oscillatory behaviour that may be present in the system due to other (classical) sources. This is where the strength of our approach lies: by contrasting spinon propagation in presence / absence of a vison, we gain access to a fundamental control measurement that uses the same Hamiltonian evolution, merely changing the initial state~\cite{Zhou_2024}. The preparation of the initial states are achieved through short subcircuits that are minimally affected by noise. This approach enables a more reliable detection of quantum coherence, as we expect similar spurious effects for the two measurements.
\begin{figure*}[!th]
    \centering
    \includegraphics[width=.98\linewidth]{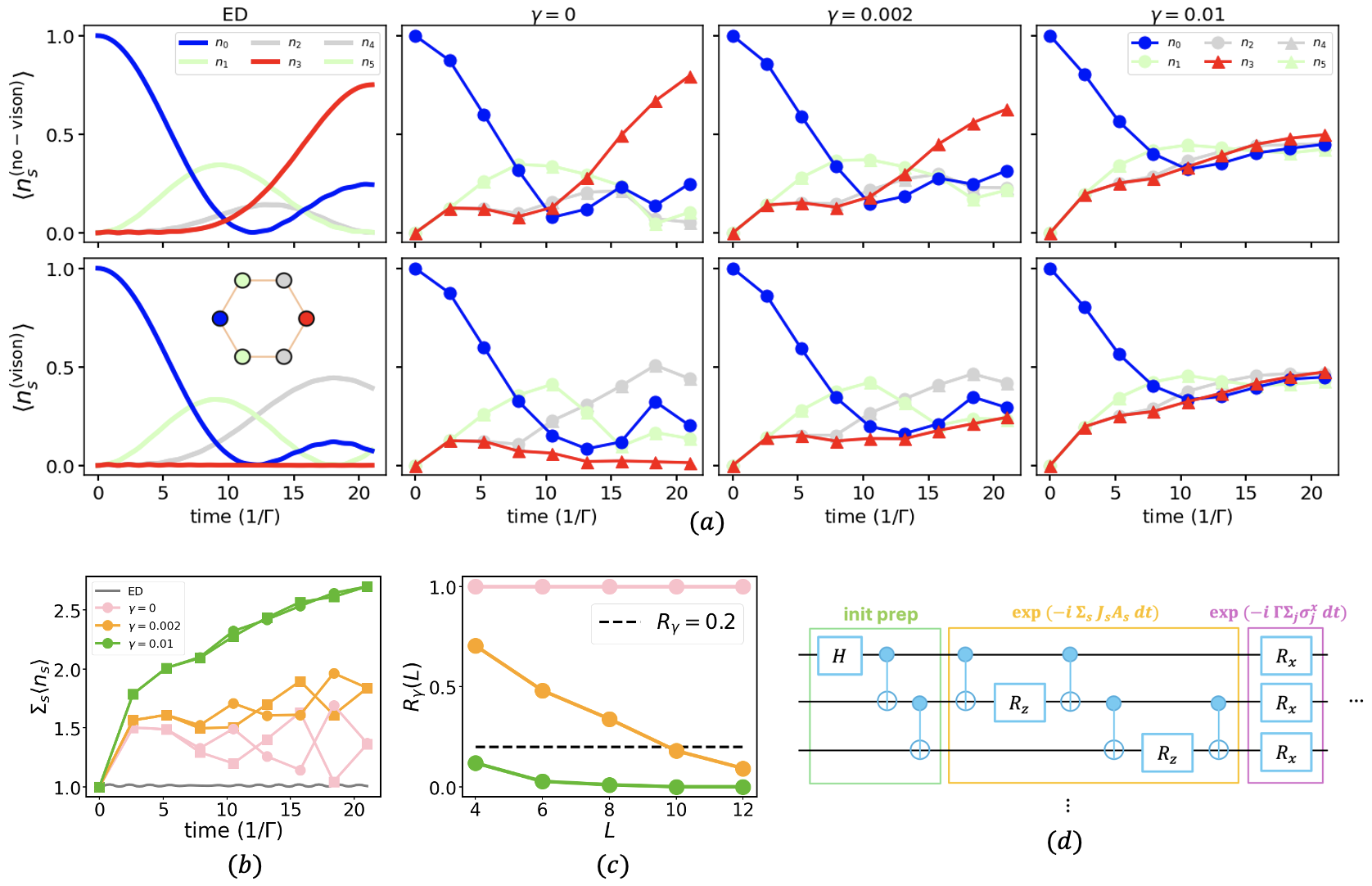}
      \caption{
      (a) Lindblad time evolution of the spinon occupation numbers $\langle n^{(\textrm{vison/no-vison})}_s \rangle$ for a ring of size $L=6$ (six sites), initialised with a single spinon at site $s=0$ and with no vison (first row) / with vison (second row) on the ring. The dramatic effect of fractional statistics ($\gamma=0$, first and second panel) survives in a crisp way when the system is coupled sufficiently weakly to a bath ($\gamma=0.002$), before becoming undetectable at stronger bath coupling ($\gamma=0.01$). The solid lines in the first panel are obtained with exact diagonalisation; the symbols correspond to optimal Trotterization, as discussed in the main text. Inset: illustration of the $L=6$ ring, with dots indicating the spinon sites, coloured according to the legend (the qubits live on the links between the dots). 
      (b) Total spinon number evolution for trotterized ED with $\gamma = 0\,, \, 0.002, \, 0.01$. The presence (solid circles) or absence (solid squares) of a vison on the ring does not affect the results). We expect all the lines with non-zero dephasing to tend to 3 spinons at large times (totally mixed state, with $1,3,5$ spinons allowed). 
      (c) $R_\gamma(L)$ as a function of ring size $L$, for $\gamma = 0, \, 0.002, \, 0.01$. With the chosen threshold at $0.2$, we see that the system has 
      Q-grade~$\simeq \infty, \, 10, \, 4$, respectively. 
      (d) Illustration of the circuit to implement such system on gate-based quantum hardware.} 
    \label{fig:lindblad}
\end{figure*}
This is illustrated in Fig.~\ref{fig:lindblad}, where we show the time evolution of the expectation value of the spinon occupation number for all the sites on a six-site ($L=6$) ring. The blockade effect introduced by the presence of a vison is dramatic ($\gamma=0$), and the contrast with / without vison remains visible for intermediate noise levels ($\gamma=0.002$), until it is heavily disrupted for strong noise ($\gamma=0.01$), and the effects of coherence and fractional statistics become difficult to detect. 
We therefore propose to use this contrast (and its disruption) to define a sensible many-body \emph{quantum coherence grade} (Q-grade) as the largest ring size where detection is sufficiently clear, for a given noise strength $\gamma$. 

In order to make this proposal concrete, we suggest the following protocol. Consider the time evolution up to the first maximum ($t_{\rm max}$~\footnote{In the single spinon approximation, tight-binding propagation gives $t_{\rm max} \, \Gamma / L \simeq 0.25$ (see SM~Note~\ref{sm_sec:particle-in-a-ring}).}) in the oscillatory behaviour of the spinon occupation number of the site $s=L/2$ diametrically opposite the initial one, $s=0$. One can then measure the difference between said occupation number at time $t_{\rm max}$ with and without the vison, normalised by the same difference in an ideal system without noise (easily obtained via classical simulations of the quantum system): 
\begin{equation}
R_\gamma(L) = 
\frac{
  \left\langle 
    n^{(\rm vison)}_{L/2}(t_{\rm max}) 
  \right\rangle_{\gamma}
  - 
  \left\langle 
    n^{(\rm no-vison)}_{L/2}(t_{\rm max}) 
  \right\rangle_{\gamma}
     }
     {
  \left\langle 
    n^{(\rm vison)}_{L/2}(t_{\rm max}) 
  \right\rangle_0
  - 
  \left\langle 
    n^{(\rm no-vison)}_{L/2}(t_{\rm max}) 
  \right\rangle_0
     }
\, , 
\label{eq:Pcoherence}
\end{equation}
where destructive interference imposes $\langle n^{(\rm vison)}_{L/2}(t_{\rm max}) \rangle_0 = 0$. 

As appropriate for gate-based quantum platforms~\cite{Byrd_2023} (and a Hamiltonian, Eq.~\eqref{eq:Ham}, that contains non-commuting terms), we implement a Trotterised time evolution~\cite{Trotter_1959} of the system to compute $R_\gamma(L)$: 
%
%
\begin{equation}
 e^{-iHt} 
 = 
 e^{-i\left(
   -\sum_s{J_s A_s} - \Gamma \sum_i{\sigma_i^x}
\right)} 
\approx 
\left[
   e^{i \frac{t}{N}\sum_s{J_s A_s}} 
   e^{i \frac{t}{N}\Gamma \sum_i{\sigma_i^x}}
\right]^N
\, ,
\label{eq:trotter}
\end{equation}
where $J_0 = -1$, and $J_s = +1$ for $s \neq 0$; $N$ is the number of Trotter steps. 
This raises a further adjustable parameter in our proposal that needs addressing. A large number of Trotter steps per unit time ensures low error in the simulated time evolution; yet, more Trotter steps extend the action of the noise (see discussion at the end of Sec.~\ref{sec:circuit}) and effectively increases $\gamma$. Here we choose empirically an optimal number of steps $N_{\rm opt} = L + 2$ as the smallest that achieves an average Trotter error $\leq 0.15$ (in the noiseless case), as discussed in SM~Note~\ref{sm_sec:circuit}. 


Our protocol is illustrated in Fig.~\ref{fig:lindblad}(c), showing the dependence of $R_\gamma(L)$ on ring size $L$ for the Lindbladian evolution in Eq.~\eqref{eq:lindblad} with $\gamma=0.002$. One clearly observes a progressive deterioration in the ability to detect the spinon-vison interference, signaling the loss of coherence in the system. We finally determine the many-body \emph{quantum coherence grade} (Q-grade) as the largest ring size $L$ where $R_\gamma(L) \geq 0.2$~\footnote{The smaller the $R_\gamma(L)$ threshold, the lower the statistical error needed to discern it accurately. To contain computational hardware costs, it is sensible and sufficient to use a relatively large threshold.}. 
%
%

\section{Circuit
\label{sec:circuit}}
In order to apply the above protocol to quantum hardware and measure the corresponding 
Q-grade, 
we need to design a suitable circuit to implement the desired Trotterised time evolution. While one could optimise this step to achieve the best possible outcome, this is not the point of our proposal. A good circuit is sufficient, so long as it is openly available and the same one can be used for different hardware platforms, to enable a direct comparison. 

Our choice of circuit implementation is illustrated in Fig.~\ref{fig:lindblad}(d). The initial state preparation (green rectangle) follows the standard GHZ state preparation protocol, given by the application of a Hadamard gate followed by a series of CNOT gates (the two different initial states are obtained by a small circuit variation, see SM). The spinon hopping term (pink rectangle) is implemented by one layer of single qubit $R_x$ gates with angle $\theta_x = 2 \Gamma dt$. The application of $ZZ$ gates, on the other hand, does allow for various constructions. One could argue that directly applying $ZZ$ is optimal as some ion-trapped quantum devices have $XX$ or $ZZ$ as native gates. However, most ion-trapped hardware only allows for a limited range of angles (usually in the range $0 \sim \pi/2$~\cite{Sorensen_1999,Debnath_2016,Maslov_2017}) for $XX$ or $ZZ$. Taking this into consideration, we propose a more generic construction given by a $R_z$ of angle $2 J_s A_s dt$ sandwiched by two CNOT gates (yellow rectangle), readily compiled by all gate-based platforms. Since the system is bi-partite, the $A_s$ term can be applied in two layers irrespective of system size. Further details can be found in SM~Note~\ref{sm_sec:circuit}. 

No variable angle is contained in the definition of the CNOT gate. As a result, choosing the standard CNOT gate as the entangling gate means that the physical or ``clock" time required for the computation scales with the number of Trotter steps, $N$. Consequently, the duration for which the noisy environment interacts with the system also scales with $N$.
%
%

\section{Devices
\label{sec:devices}}
\begin{figure*}[!th]
    \centering    
\includegraphics[width=.98\linewidth]{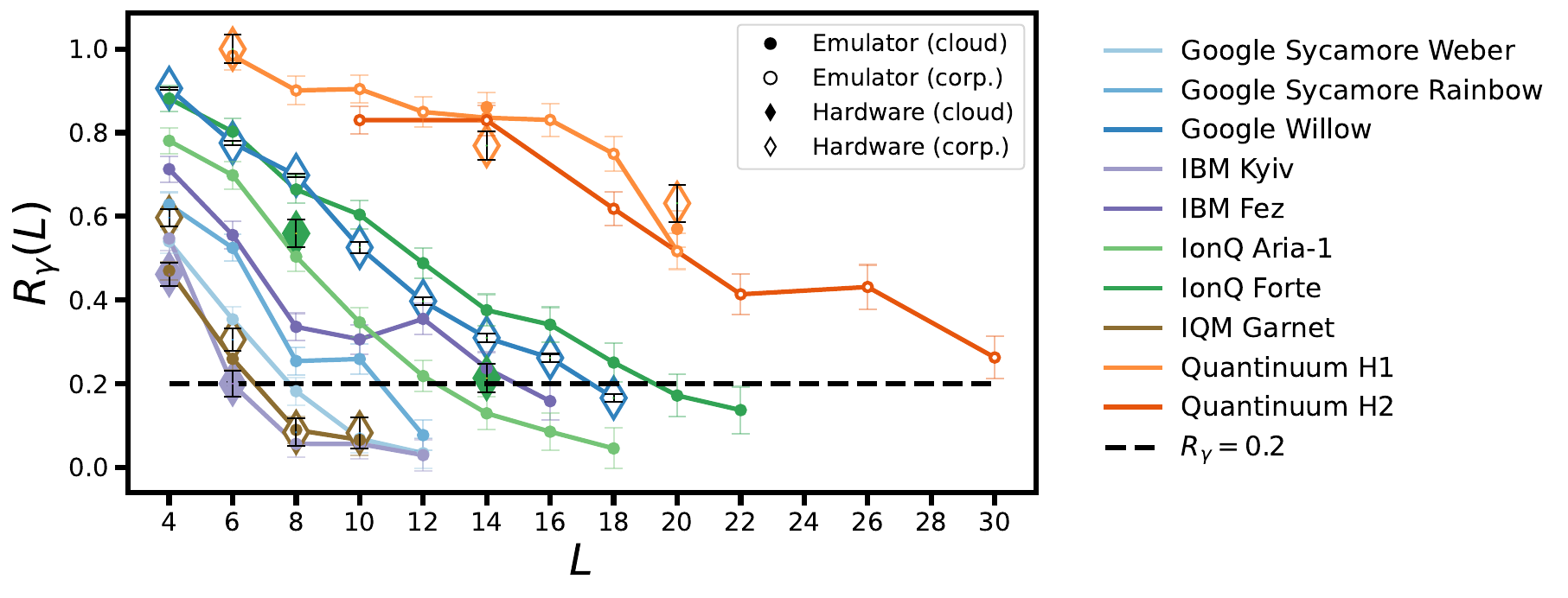}
      \caption{Coherence measure $R_\gamma(L)$ as a function of ring size $L$, for different architectures:
      Google Sycamore Weber, 
      Google Sycamore Rainbow, 
      Google Willow, 
      IBM Kyiv, 
      IBM Fez, 
      IonQ Aria-1, 
      IonQ Forte, 
      IQM Garnet, 
       Quantinuum H1, and Quantinuum H2. 
      Solid dots are emulator results ($1000$ shots for $L\leq16$ and $2000$ for $L>16$), and soild diamonds are hardware results ($1000$ shots), that we ran directly through cloud platforms. The computation of statistical uncertainty is discussed in detail in the SM Note~\ref{sm_sec:stat_error}. Empty circles and empty diamonds are emulator and hardware results that were kindly provided by the respective platform teams 
      (see SM Note~\ref{sm_sec:hardware} for details).
      The threshold at $0.2$ identifies the many-body quantum coherence grade (Q-grade) for each architecture.} 
    \label{fig:hardware}
\end{figure*}
We are finally in the position to investigate the 
Q-grade 
for different gate-based quantum devices: Google~\cite{google_quantum_ai}, IBM~\cite{ibm}, IonQ~\cite{ionq}, IQM~\cite{iqm}, and Quantinuum~\cite{quantinuum}. This can be done directly using the quantum hardware, or via a suitable emulator that is capable of capturing the corresponding noise with quantitative accuracy. 

The central result of this work is shown in Fig.~\ref{fig:hardware}, where we plot the coherence measure $R_\gamma(L)$ in Eq.~\eqref{eq:Pcoherence} as a function of $L$, for different hardware (or emulators thereof)~\footnote{
Details of how these data were extracted for each architecture are given in the SM~Note~\ref{sm_sec:hardware}.
}. 

The crossing points with the chosen threshold $0.2$ give then the respective 
Q-grades. 
While one should not assign too specific a meaning to the actual Q-grade value, its importance lies in the comparison it allows -- most importantly in the ability it provides to track the evolution of many-body coherence capabilities of quantum hardware as it evolves and improves: from Google Sycamore Weber to Rainbow to Willow; from IBM Kyiv to Fez; from IonQ Aria-1 to Forte; from Quantinuum H1 to H2. Through the years, progress can be tracked and the possible appearance of a many-body quantum coherence Moore's law can be investigated. 
%
%

\section{conclusions
\label{sec:conclusions}}
We showed how to define a many-body quantum coherence length scale -- the Q-grade -- via anyon interference in a simple Ising-chain set up. We demonstrated how this can be readily implemented in current gate-based quantum hardware devices, including Google, IBM, IonQ, IQM, and Quantinuum. Our work highlights the potential of establishing standardized quantum tests to assess and compare available hardware while also providing a crucial benchmark for tracking progress as quantum technologies evolve. The Q-grade data we present underscore the progress made across all quantum platforms with each new hardware iteration. 
Our proposal provides a standardized measure of many-body coherence; this is a vital aspect of the overall capability of a quantum device, along with the number of qubits, connectivity, and scalability. 

The fundamentals of anyon interference underlying the definition of Q-grade are independent of the current state-of-the-art in hardware development. 
We intend our work to inform the creation of a live web interface that the platform teams can access and update directly, where the latest developments and achievements can be demonstrated, and where progress on available quantum coherence resources can be logged and monitored over time. The willingness of teams across various quantum platforms to contribute their data to this project strongly suggests the potential for a broad, community-wide adoption. Recording progress in the live web interface will enable tracking the growth (\`a la Moore's law) of coherence in gate-based quantum platforms, and allow more informed predictions about future quantum capabilities. 
%
%

\section*{Acknowledgements}
Y.T. would like to thank M.~Rutter and H.~Sanghera for their support with scientific and quantum computing software. 
We are grateful to E.~D.~Dahl and F.~Tripier at IonQ for insightful and helpful discussions, and to IBM for their systems support. 
We are grateful to S.-H.~Lin, M.~Iqbal and H.~Dreyer at Quantinuum for providing us with H1 and H2 emulator and hardware results. 
We would like to thank S.~Kumar, E.~Rosenberg, and P.~Roushan from the Google Quantum AI team for providing hardware results on the Willow processor, and for insightful discussions in particular about the error bars in Fig.~\ref{fig:hardware}. 
We would like to thank A.~Calzona and M.~J.~Thapa at IQM for insightful discussions and for providing access, instructions and data. 
This work was funded in part by the Engineering and Physical Sciences Research Council (EPSRC) grants No.~EP/T028580/1 and No.~EP/V062654/1 (C.C.), No.~EP/W005484(O.S.), by NSF Grant No. DMR-1945395 (A.R.), and by DOE Grant No.~DE-FG02-06ER46316 (C.Ch.). Research at Perimeter Institute is supported in part by the Government of Canada through the Department of Innovation, Science and Industry Canada and by the Province of Ontario through the Ministry of Colleges and Universities.
%
%

\putbib
\end{bibunit}
%
%

\clearpage
\newpage

\appendix

\setcounter{figure}{0}
\renewcommand{\thefigure}{S\arabic{figure}}
\renewcommand{\appendixname}{SM Note}

\begin{bibunit}

\onecolumngrid
\vspace{\columnsep}
\begin{center}
{\Large\bf Supplementary Material for:} 

\vspace{0.2 cm}
{\large\bf ``Standardized test of many-body coherence in gate-based quantum platforms''}

\vspace{0.5 cm}
Yi Teng,$^{1}$
Orazio Scarlatella,$^{1}$
Shiyu Zhou,$^{2}$
Armin Rahmani,$^{3}$
Claudio Chamon,$^{4}$
and 
Claudio Castelnovo,$^{1}$ 

\vspace{0.3 cm}

$^{1}$TCM Group, Cavendish Laboratory, University of Cambridge, Cambridge CB3 0HE, UK

$^{2}$Perimeter Institute for Theoretical Physics, Waterloo, Ontario, Canada N2L 2Y5

$^{3}$Department of Physics and Astronomy and Advanced Materials Science and Engineering Center, Western Washington University, Bellingham, Washington 98225, USA

$^{4}$Department of Physics, Boston University, Boston, MA, 02215, USA

\end{center}
\vspace{\columnsep}
\twocolumngrid
%
%

\section{Model
\label{sm_sec:model}}
\begin{figure}[h]
    \centering
    \includegraphics[width=.45\textwidth]{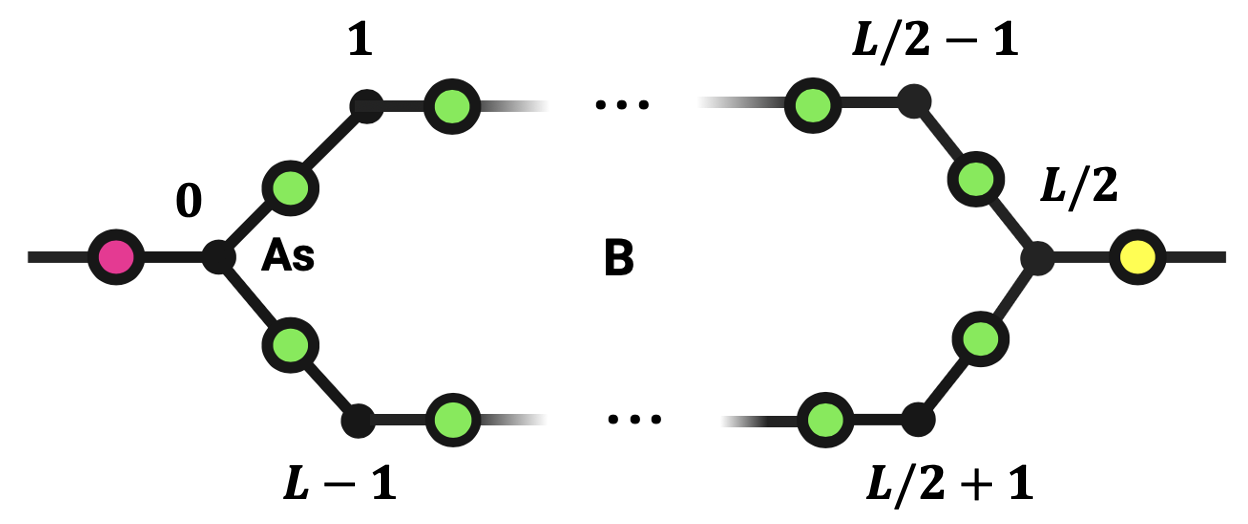}
      \caption{Schematic illustration of the system considered in this work. The colour-filled large dots represent the qubits, whereas the the solid dots represent spinon sites ($s=0,\ldots,L-1$). Spinon operators are associated with the product of $z$ components of two adjacent qubits, except for the leftmost and rightmost sites, $s=0,L/2$, that involve three qubits. 
      The leftmost (pink) and rightmost (yellow) qubits are fixed (see text). The vison (or plaquette) operator is given by the product of the $x$ components of the qubits around the ring (green).} 
    \label{fig:system_app}
\end{figure}

In the main text, we framed the discussion of the interferometer geometry in terms of a 1D Ising chain with twisted boundary conditions. While this approach is concise and highlights the essence of the interferometer, it conceals the connection to Kitaev's toric code, from which the roles of the spinon and the vison in giving the interference patterns is most clear. The connection to Kitaev's model can be made using the graph shown in Fig.~\ref{fig:system_app}, in which we further add two qubits on the flanks of the leftmost and rightmost spinon sites on the ring in Fig.~\ref{fig:system} of the main text.
Correspondingly, we modify the operators $A_{s=0,L/2}$ to be the product of the \emph{three} adjacent $\sigma_i^z$, while keeping all other $A_{s \neq 0,L/2}$ as the products of the \emph{two} adjacent $\sigma_i^z$. All these star operators commute with the plaquette operator given by the product of the $L$ $\sigma_i^x$ around the ring, $B = \prod_{\rm ring} \sigma_i^x$. 
One can then recognise the system as being a version of Kitaev's toric code~\cite{Kitaev_2003}, one that is defined on the graph with a ring of stars labeled by $s=0,\dots, L$ and a single plaquette (no label needed), whose Hamiltonian reads 
\begin{equation}
H = - \lambda_A \sum_s \;A_s 
- \lambda_B \;B 
\qquad 
(\lambda_A, \lambda_B > 0)
\label{eq:HamTC}
\, . 
\end{equation}
Both star $A_s$ and plaquette $B$ operators have eigenvalues $-1$ or $+1$, which corresponds to the presence or absence of an `e' (spinon) or an `m' (vison) quasiparticle in the system, which have mutual semionic statistics. In our model, the vison can only be located on the ring, whereas the spinons can live on any site $s$ of the ring (solid black dots in Fig.~\ref{fig:system_app}). 

We are specifically interested in the behaviour of a single spinon, in presence/absence of a vison on the plaquette. For this reason, it is convenient to fix the $z$-component of the leftmost and rightmost qubits to point in opposite directions. This enforces an odd spinon number in the system, and consequently the lowest energy sector is that containing a single spinon, rather than the vacuum. Due to their mutual semionic statistics, the presence of a vison causes perfect destructive interference between the world-lines of the spinon moving, say, from one side to the other of the ring, along the upper vs. lower leg. This means that a spinon initialised on the leftmost site of the ring cannot reach the rightmost site if a vison is present, barring decoherence due to noise. 

Our system in the main text corresponds to the case where the leftmost qubit is projected onto the $\sigma^z=-1$ state while the rightmost qubit is projected onto the $\sigma^z=+1$ state, and we set $\lambda_A=J$ and $\lambda_B=0$, as well as introduce a uniform transverse field $- {\Gamma} \sum_i\sigma_i^x$ (see Eq.~\eqref{eq:Ham} in the main text). 

The mutual semionic statistics of a spinon and a vison can be illustrated, in the case of a $L = 6$ (hexagonal) ring, by considering the processes that move a spinon, initialised on the leftmost site $s=0$, around the ring to the rightmost site $s=L/2$, along the top and bottom leg of the ring, respectively: 
\begin{eqnarray}
   \mathcal{P}_{\rm top} &=& \Gamma^3 \,\sigma_1^x \sigma_2^x \sigma_3^x
   \\ 
   &\,& \nonumber\\
   \mathcal{P}_{\rm bottom} &=& \Gamma^3 \,\sigma_4^x \sigma_5^x \sigma_6^x
   \\
   &=& 
   \Gamma^3 \,\sigma_4^x \sigma_5^x \sigma_6^x 
   \,(\sigma_1^x \sigma_2^x \sigma_3^x 
   \sigma_1^x \sigma_2^x \sigma_3^x)
   \nonumber \\ 
   &=& B \, \Gamma^3\,  \sigma_1^x \sigma_2^x \sigma_3^x 
   \nonumber \\ 
   &=& B \,\mathcal{P}_{\rm top}
\, . 
\nonumber 
\end{eqnarray}
We see explicitly that in presence of a vison, $B = -1$, the two pathways have opposite sign and interfere destructively, and the spinon is strictly prevented from visiting the site $s=L/2$; in contrast, without the vison, $B = +1$, the two pathways add constructively, and the spinon propagates freely across the ring. 
This is analogous to Aharonov-Bohm blockade of an electron charge around a magnetic $\pi$ flux~\cite{Aharonov_1959}, except that in this case the flux (vison or `m' quasiparticle) is emergent and borne out of the same qubits from which the charge (`e' or spinon quasiparticle) emerges. 
%
%

\section{Decoherence
\label{sm_sec:dephasing}}
In an ideal system, semionic statistics leads to perfect destructive interference and complete spinon blockade. In real systems, noise-induced decoherence is expected to gradually enable leakage of the spinon across the ring, even when a vison is initially present. On general grounds, decoherence causes both the statistical angle to become ill-defined, namely the vison number to decay, and the spinon number to relax. 

To model the effects of noise we consider an open system where each qubit is coupled to an isotropic bath (a choice that is both simple and also proved quantitatively accurate to model quantum circuits~\cite{Zhou_2024}). 
The evolution of the density matrix is governed by the Lindblad master equation~\cite{Lindblad_1976}:
\begin{equation}
\dot{\rho} 
= 
- i [H,\rho] 
+ \gamma \sum_i \left(
  \sigma^x_i \rho \sigma^x_i + \sigma^y_i \rho \sigma^y_i + \sigma^z_i \rho \sigma^z_i - 3 \rho
\right) 
\, ,
\label{eq:isotropic dephasing}
\end{equation}
where $i$ runs over all the qubits (apart from the two fixed ones). Note that with such dephasing noise, the system inevitably relaxes to the maximally mixed state (i.e., a density matrix proportional to the identity matrix) at large times. 

\begin{figure}[h]
    \centering
    \includegraphics[width=.3\textwidth]{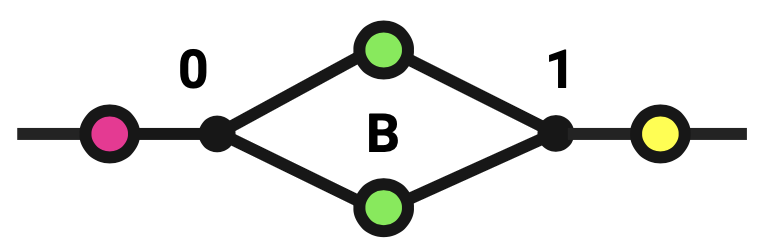}
      \caption{Illustration of the $L=2$ two-qubit system, following Fig.~\ref{fig:system_app}. Again, qubits in pink and yellow are fixed, and the two spinon sites are labelled $0$ and $1$.} 
    \label{fig:4spins}
\end{figure}

The simplest system that exhibits the interference effect of interest is one with two qubits on the ring, illustrated in Fig.~\ref{fig:4spins}. 
Note that in this case, with fixed external qubits that enforce an odd number of spinons, only the single-spinon sector is allowed and thus the total spinon number cannot fluctuate. The Hilbert space is spanned by four states associated with the two qubits on the ring ($\sigma^z=\pm 1$). It is however more convenient to work in the eigenbasis of the spinon, $A_{s=0,1} = \pm 1$, and vison, $B = \pm 1$, operators. We shall label these states as: 
$0$, spinon at $s=0$ and no vison; 
$0'$, spinon at $s=0$ and vison; 
$1$, spinon at $s=1$ and no vison; 
and $1'$, spinon at $s=1$ and vison. 

The time evolution equation under isotropic single-qubit depolarization, Eq.~\eqref{eq:isotropic dephasing}, can be solved analytically, and the elements of the density matrix in the spinon and vison site occupation basis can be written as: 
\begin{eqnarray}
\rho_{00} &=& \frac{1}{4} + C_{1} e^{-8\gamma t}
\label{eq:solution_n=1}
\\
\rho_{11} &=& \frac{1}{4} + C_{2} e^{-8\gamma t}
\nonumber \\ 
\rho_{0'0'} &=& \frac{1}{2} + e^{-8\gamma t} [C_{3} + C_{4} \cos(4\Gamma t) + C_{5} \sin(4\Gamma t)]
\nonumber \\ 
\rho_{1'1'} &=& \frac{1}{2} + e^{-8\gamma t} [C_{3} - C_{4} \cos(4\Gamma t) - C_{5} \sin(4\Gamma t)]
\, . 
\nonumber 
\end{eqnarray}
where the $C_{i}$ are constants fixed by the initial conditions; all other density matrix elements vanish. Note that if we initialise the system with the spinon on the left in the presence of a vison on the ring ($\rho_{00} = 1$ and all other elements set to $0$ at $t = 0$), we obtain: 
%
\begin{eqnarray}
\rho_{00} &=& \frac{1}{4} + \frac{3}{4} e^{-8\gamma t}
\, , 
\quad
\rho_{11} = \frac{1}{4} - \frac{1}{4} e^{-8\gamma t}
\nonumber \\ 
\rho_{0'0'} &=& \frac{1}{4} - \frac{1}{4} e^{-8\gamma t}
\, , 
\quad
\rho_{1'1'} = \frac{1}{4} - \frac{1}{4} e^{-8\gamma t}
\, .
\label{eq:isotropic sol1}
\end{eqnarray}
If we initialise the state with a spinon on the left in the absence of a vison ($\rho_{0'0'} = 1$ and all others $0$ at $t=0$), then the solution is: 
\begin{eqnarray}
\rho_{00} &=& \frac{1}{4} - \frac{1}{4} e^{-8\gamma t}
\label{eq:isotropic sol2}
\\
\rho_{11} &=& \frac{1}{4} - \frac{1}{4} e^{-8\gamma t} 
\nonumber \\ 
\rho_{0'0'} &=& \frac{1}{4} + \frac{1}{4} e^{-8\gamma t} + \frac{1}{2}e^{-8\gamma t} \cos(4\Gamma t)
\nonumber \\
\rho_{1'1'} &=& \frac{1}{4} + \frac{1}{4} e^{-8\gamma t} - \frac{1}{2}e^{-8\gamma t} \cos(4\Gamma t) 
\, .
\nonumber 
\end{eqnarray}

The left and right spinon occupation numbers are given by $\rho_{00} + \rho_{0'0'}$ and $\rho_{11} + \rho_{1'1'}$, respectively. 
\begin{figure}[!ht]
\centering
\includegraphics[width=.48\textwidth]{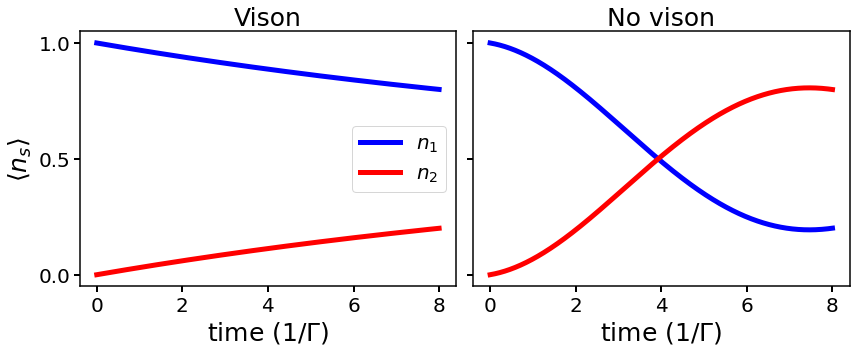}
\caption{Analytical time evolution of the spinon site occupation numbers for the two-qubit ring system ($L=2$) with $J=1$, $\Gamma=0.1$, and isotropic dephasing strength $\gamma = 0.008$.} 
\label{fig:2_site_results}
\end{figure}
The resulting behaviour is illustrated in Fig.~\ref{fig:2_site_results}, to be contrasted with Fig.~\ref{fig:lindblad} in the main text. 

Notice that the two-qubit ($L=2$) ring system can be simulated without Trotterisation (the Hamiltonian is merely $\sum_i\sigma^x$). When implemented on a circuit, the noise is only due to the imperfect single gate fidelity, which can be very high and one does not observe any decay out to very large time scales. 
%
%

\section{\label{sm_sec:particle-in-a-ring}
A particle in tight-binding ring with/without flux}
Consider a one-dimensional closed ring, composed of $L$ (even) sites labeled by 
$x\in \{0,1,2,\dots, L-1\}$. 
Neighboring sites are connected by a hopping amplitude $\Gamma$. The eigenenergies of a single particle hopping in this system are given by $\epsilon_k = -2\Gamma\cos k$, with $k$ taking discrete values 
$k_j=\frac{2\pi}{L}(j+\frac{\phi}{2\pi})$, where $\phi$ is the flux through the ring and 
$j\in\{-L/2,-L/2+1,\dots, L/2-1\}$. 
The wavefunction at site $x$ and time $t$, if the particle is launched from $x=0$ at $t=0$, is 
\begin{align}
\label{eq:ring-wavefunction}
    \Psi_\phi(x,t) = \frac{1}{L}\;\sum_{j=-L/2}^{L/2-1} 
    \; e^{-i\,t\,2\,\Gamma \cos k_j}
    \;e^{i\, k_j\, x}
    \, . 
\end{align}
For flux $\phi=\pi$ the wavefunction obeys anti-periodic boundary conditions, 
$\Psi_\pi(x,t) = - \Psi_\pi(x+L,t)$. In this case the 
$L$ 
quantized values $k_j$ are symmetric with respect to the origin, and exchanging $k\to -k$ in Eq.~\eqref{eq:ring-wavefunction} yields $\Psi_\pi(x,t) = \Psi_\pi(-x,t)$. In particular, 
we have 
$\Psi_\pi(L/2,t) = \Psi_\pi(-L/2,t)=-\Psi_\pi(L/2,t)$, and thus $\Psi_\pi(L/2,t)=0$ 
at all times.
\begin{figure}
    \centering
    \includegraphics[width=.45\textwidth]{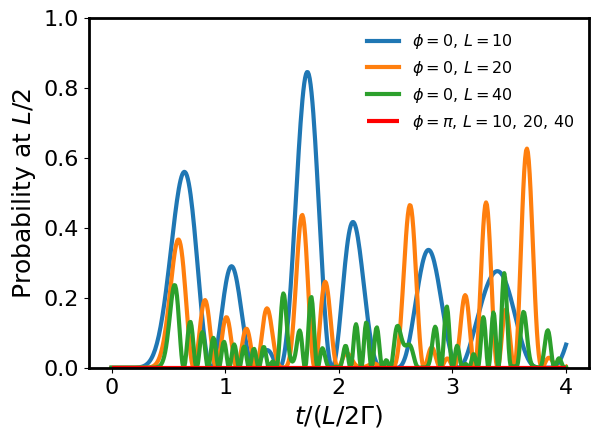}
    \caption{Probability that a particle launched for $x=0$ in a tight-binding ring of length $L$ reaches the mid-point at $x=L/2$, for the cases when a flux $\phi=0,\pi$ threads the ring. The particle first reaches $x=L/2$ at time $t \approx L/4\Gamma$. Notice that subsequent probability peaks signal revivals at later times.}
    \label{fig:revivals}
\end{figure}

For flux $\phi=0$ the wavefunction obeys periodic boundary conditions, 
$\Psi_0(x,t)=\Psi_0(x+L,t)$. 
The probability $p_0(x,t)= |\Psi_0(x,t)|^2$ can be computed numerically using Eq.~\eqref{eq:ring-wavefunction}. In particular, 
$p_0(L/2,t)$ in Fig.~\ref{fig:revivals} shows a peak 
for 
$\Gamma t / T \approx 1/4$, 
as well as later revivals (e.g., a notable one for $\Gamma t / L \approx 3/4$). 

Note that the our Q-grade tests the many-body quantum coherence of a system both in time and in number of qubits, distinguishing the flux/no flux cases by focusing on the time of first arrival of the particle at site labeled $s=L/2$, which stand the farthest from the launch site $s=0$. There are a number of recurrences of the probability for the particle to be detected at site $s=L/2$. These later recurrences could be used in the future to devise different grades that probe the same number of qubits for longer and longer times, therefore testing many-body quantum coherence times with fixed number of qubits. 
%
%

\section{Initial state preparation and Trotterised evolution in quantum circuits
\label{sm_sec:circuit}}
Let us discuss here details of the choice of $t_{\rm max}$ and optimal Trotter step, as well as details of the overall quantum circuit implementation. 
%
%

\subsection{Optimal evolution time and Trotter step}
\begin{figure}[h]
    \centering
    \includegraphics[width=.4\textwidth]{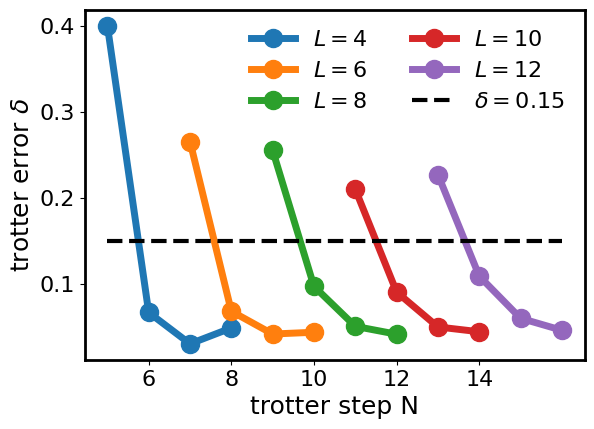}
      \caption{Trotter error, defined in Eq.~\eqref{eq:trotter error}, for different system sizes as a function of number of Trotter steps. We empirically choose a threshold error $\delta = 0.15$ to find a corresponding optimal trotter step of $N_{\rm opt} = L+2$.} 
    \label{fig:trot_err}
\end{figure}
We choose the optimal evolution time $t_{\rm max}$ to be the first peak in spinon density at site $L$. In other words, the point where the contrast between the cases with and without a vison becomes most pronounced (for the first time). This $t_{\rm max}$ is obtained from circuit evolution with a large number of Trotter steps and is rounded to the nearest integer value. We found $t_{\rm max} = 16, 21, 27, 32, 38, 43, 48, 54, 59, 65$ for $L = 4, 6, 8, 10, 12, 14, 16, 18, 20, 22$. This sequence can roughly be extrapolated as $t_{\rm max} \simeq 5 + 2.25 L$, scaling linearly with $L$ as expected from the quantum Ising chain considerations in SM Note~\ref{sm_sec:particle-in-a-ring}. Given $t_{\rm max}$, the optimal Trotter step is set by the minimal number of steps needed to reach a certain level of accuracy in describing the time evolution of the system up to time $t_{\rm max}$. Quantitatively, we use the following formula to calculate an average Trotter error:
\begin{equation}
\delta 
= 
\sqrt{ \frac{1}{2 N L} \sum_{\textrm{v, /v}} \sum_{s,t} \left[
\langle n^{\rm ED}_s(t) \rangle 
- 
\langle n^{\rm Trotter}_s(t) \rangle
\right]^2}
\, , 
\label{eq:trotter error}
\end{equation}
where $s=0,\ldots,L-1$, and $t=0,\ldots,t_{\rm max}$ denote sites and time steps, `v, /v' represents summing over the vison and no vison cases, $n_s$ is the spinon density operator, and $N$ stands for the total number of Trotter steps (the normalisation factor is chosen so that $\delta \leq 1$). We choose $\delta = 0.15$ as our threshold, and we empirically find an optimal trotter step $N_{\rm opt} =  6, 8, 10, 12, 14$ for $L = 4, 6, 8, 10, 12$. We extrapolate this to an optimal Trotter step $N_{\rm opt} = L+2$, where $L$ is the size of the ring (see Fig.~\ref{fig:trot_err}). 
%
%

\subsection{Quantum circuit details}
\begin{figure*}[!th]
    \centering
    \includegraphics[width=.8\linewidth]{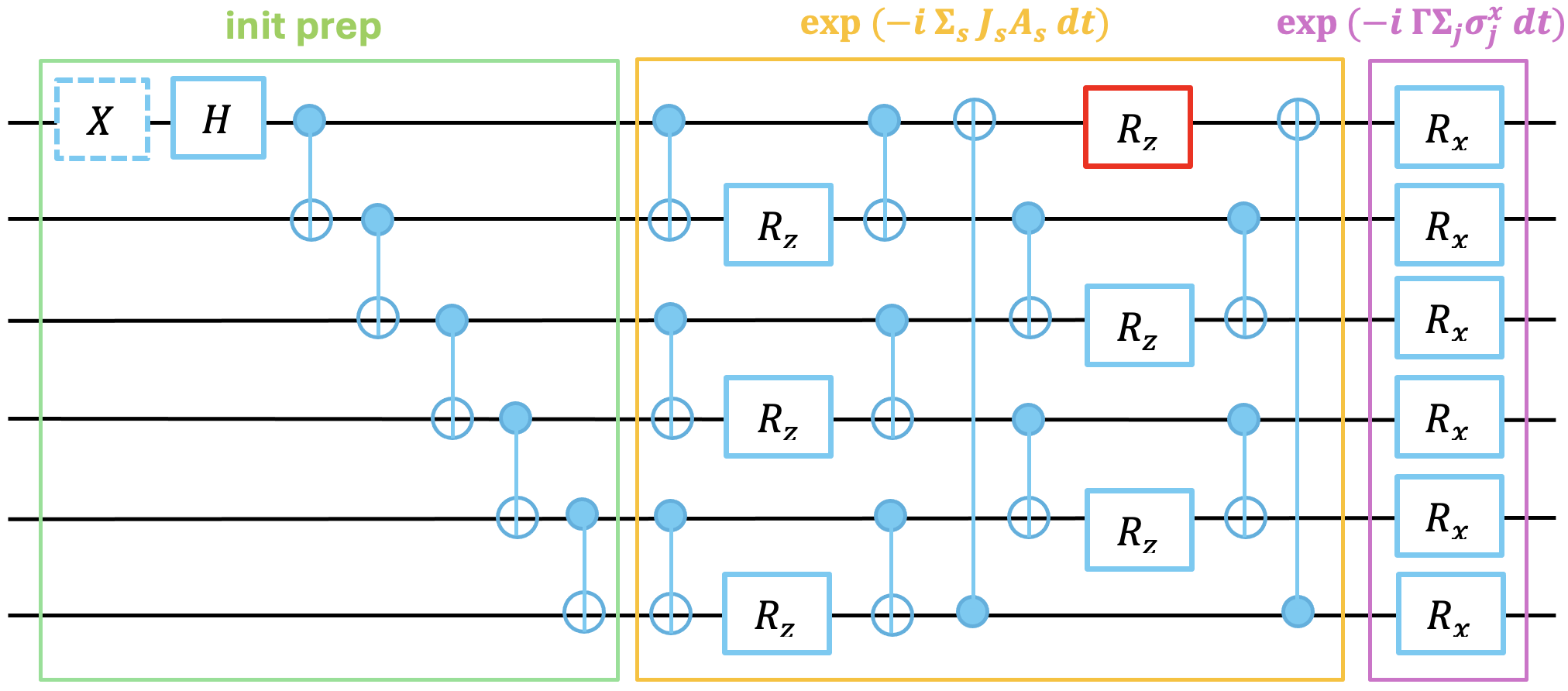}
      \caption{Full circuit used for the Trotterized evolution ($L = 6$ for illustration). The initial state preparation is shown in the green rectangle. We apply a Hadamard gate on the first qubit, followed by consecutive CNOT applications on $(i, i+1)$ qubit, $i = 1,2,3,...,L-1$ entangling all the qubits. An $X$ gate (blue dashed rectangle) is (optionally) applied between the Hadamard and the first CNOT gate to introduce a vison on the ring. The parts enclosed in the orange and pink rectangle show one Trotter step of the Hamiltonian evolution. The CNOT + Rz($\theta_z$) + CNOT combination effectively executes $\exp(i\theta_z\, ZZ / 2)$. In the first layer, this combination is applied to $(i, i+1)$ for $i = 1,3,5,...,L-1$. In the second layer, it is applied to $(i,i+1)$ for $i = 2,4,6,...,L-2$ and to $(L, 1)$. Most $R_z$ gates have an angle of $\theta_z = 2J \,t_{\rm max} / N_{\rm opt}$, except the one in red which has an angle of $-\theta_z$ due to the minus sign in $J_0$. $R_x$ with an angle $\theta_x = 2\Gamma \,t_{\rm max} / N_{\rm opt} $ executes the transverse field. The Trotter step (orange and pink rectangle sequence) is then repeated $N_{\rm opt}$ times.}
    \label{fig:full_circuit}
\end{figure*}
The quantum circuit is shown in Fig.~\ref{fig:full_circuit}. The initialisation follows a standard GHZ state preparation. We did not use more efficient protocols with logarithmic growth of circuit depth~\cite{Cruz_2019}, as it requires connectivity beyond nearest neighbors; this is done in purpose to conform to most hardware topologies. One Trotter step is shown within the orange and pink rectangles, with angle parameters for $R_z$ and $R_x$ rotations equal to $\theta_z = 2J \,t_{\rm max} / N_{\rm opt}$ and $\theta_x = 2\Gamma \,t_{\rm max} / N_{\rm opt} $, except that the angle for the $R_z$ gate depicted in red has an additional minus sign. 
The pseudocode for circuit implementation is given in the caption of Fig.~\ref{fig:full_circuit}. 
%
%

\section{\label{sm_sec:hardware}
Emulator and hardware implementation details}
In this section we present the protocols with which we obtained the emulator and/or hardware results presented in Fig.~\ref{fig:hardware} in the main text. 
%
%

\subsection{Google}
We used the \texttt{google\_cirq} library to extract the median device calibration data and build the corresponding noise model for Google Sycamore Weber and Google Sycamore Rainbow. Physical qubits are then picked by hand for each size $L$ and are translated to the target circuit using the same library. We prioritize good two-qubit gate fidelity during the qubit selection process (information about the exact qubits picked is available upon request). Before using the emulator, one needs to pick a gateset in which the original circuit is translated into. The three options available from \texttt{google\_cirq} are SqrtIswap, Sycamore and CZ gatesets. In practice, we find that the CZ gateset gives the best results, and we adopt it throughout. Finally, \texttt{qsimcirq} is used to execute the simulations and to obtain the emulator data for Weber and Rainbow. 
Willow hardware results were kindly provided by Google Quantum AI. The circuits were compiled into CPhase gates and arbitrary-angle single-qubit gates. The raw circuits were processed using gauge compiling around the CPhase gates, with single-qubit gates merged into phased XZ gates, followed by dynamical decoupling. For each data point, 100 random instances were used to calculate the mean for circuits with and without vison. 10,000 shots were taken for each circuit. 
%
%

\subsection{IBM}
We used the \texttt{AerSimulator} in the \texttt{qiskit\_aer} library and set \texttt{optimization\_level} (circuit compilation optimization) to 1 in \texttt{generate\_preset\_pass\_manager}. Next,  \texttt{EstimatorV2} in \texttt{qiskit\_ibm\_runtime} is used to build a noisy emulator from \texttt{AerSimulator}. For the hardware simulation, further parameters in \texttt{EstimatorOptions} are set as follows: resilience level to 1, optimization level to 0 (the circuit has been optimized by \texttt{generate\_preset\_pass\_manager}), dynamical decoupling enable to True, and dynamical decoupling sequence type to XY4. 

We note the pronounced shoulder in the results shown in Fig.~\ref{fig:hardware} in the main text for the IBM hardware around $L=12$, which we attribute to the fact that their architecture features qubit loops of length $12$ and is therefore particularly well-suited to implement this system size.
%
%

\subsection{IonQ}
We accessed IonQ emulators Aria-1 and Forte 
via Microsoft Azure Quantum (\texttt{azure.quantum}). 
Details of the noise emulator are given in their online documentation~\cite{ionq_noise_model}. The option parameters in this case do not appear to make a noticeable difference. In particular, the debias option is not actually executed in the emulator. 

We additionally accessed IonQ Forte hardware via AWS Braket. The device initialization and circuit compilation are all completed in an AWS notebook with no further parameters chosen by us, i.e., we used the default setting a user experiences when accessing IonQ Forte via the AWS Cloud Service. Note that debiasing is automatically disabled on AWS for less than 2500 shots. 
%
%

\subsection{IQM}
We used the \texttt{IQMFakeApollo} in \texttt{qiskit\_iqm} to simulate IQM Garnet. The circuit layouts were optimized using Qiskit’s transpiler (optimization level = 1) and mapomatic package. All Z rotations are implemented virtually. Hardware data were kindly provided by Manish Thapa and Alessio Calzona. Both dynamical decoupling (DD) and readout error mitigation were applied in hardware runs. Our DD involves a combination of XX, XYXY and XYXYYXYX sequences. These sequences are automatically placed by looking at idles which have fitting sizes. For example, if the idles are short (i.e., 3 times the duration of X gate), XX sequence is symmetrically placed. If the idles are longer than 8 times the duration of X gate, XYXYYXYX sequence is placed accordingly. 
%
%

\subsection{Quantinuum}
H1 hardware and emulator results for Quantinuum were kindly provided by Sheng-Hsuan Lin, Mohsin Iqbal and Henrik Dreyer.
The highest optimization level (level 2) in their library \texttt{pytket} was used for circuit compilation, whereby the CNOT-Rz-CNOT series is converted to a ZZ rotation, reducing the circuit depth significantly. 
We independently accessed the Quantinuum emulator via Microsoft Azure Quantum (\texttt{azure.quantum}) for selected ring sizes ($L=7,10$) and obtained results in quantitative agreement with the ones shown in Fig.~\ref{fig:hardware} in the main text. 
%
%

\section{\label{sm_sec:stat_error}
Statistical uncertainty of the Q-grade}
The Q-grade is defined in Eq.~\eqref{eq:Pcoherence} in the main text. We simplify the notation and write it as follows
\begin{equation}
R_{\gamma} = 
\frac{
    n^{\rm v}_{\gamma} 
  - 
  n^{\rm nov}_{\gamma} 
     }
     {
    n^{\rm v}_{0} 
  - 
  n^{\rm nov}_{0} 
     }
\, , 
\label{eq:Pcoherence_simp}
\end{equation}
where $n$ denotes the expectation value of the spinon occupation number on the rightmost site, $v$ and $nov$ represent the different initialisation with and without vison, $\gamma$ and $0$ stand for the noisy and ideal cases. The spinon occupation number for the ideal case, $n^{\rm v}_0$ and $n^{\rm nov}_0$ are obtained to significantly higher accuracy compared to their noisy counterparts, so we omit their contributions to the uncertainty and follow the statistical error propagation formula 
\begin{equation}
\Delta R_{\gamma} = \sqrt{
\left(
\frac{\partial R_{\gamma}}
{\partial n^{\rm v}_{\gamma}}
\right)^2 
(\Delta n^{\rm v}_{\gamma})^2 
+ 
\left(
\frac{\partial R_{\gamma}}{\partial n^{\rm nov}_{\gamma}}
\right)^2 
(\Delta n^{\rm nov}_{\gamma})^2
}
\, .
\label{eq:stat_error_prop}
\end{equation}

Since the spinon occupation numbers ($n^{\rm nov}_{\gamma}, n^{\rm v}_{\gamma}$) are probability-like variables ranging between 0 and 1, it is sensible to assume that they follow a binomial distribution and have statistical error $\Delta n = \sqrt{n(1-n)/N_{\rm shots}}$ with $N_{\rm shots}$ being the number of shots. We then arrive at the final error formula  
\begin{equation}
\Delta R_{\gamma} =  \sqrt{\frac{(1 - n^{\rm v}_{\gamma}) n^{\rm v}_{\gamma} + (1 -  n^{\rm nov}_{\gamma}) n^{\rm nov}_{\gamma}}{N_{\rm shots}(n^{\rm v}_{0} -  n^{\rm nov}_{0})^2}}
\, .
\label{eq:stat_error_formula}
\end{equation}

With Eq.~\eqref{eq:stat_error_formula}, we find the statistical uncertainty and present the error bars in Fig.~\ref{fig:hardware}. Note that this only takes into account the statistical error, while other error sources, such as readout errors or memory errors, could result in a larger uncertainty.
%
%

\putbib
\end{bibunit}


\end{document}